\documentclass[twocolumn,tightenlines,superscriptaddress,showpacs,amsmath,amssymb,aps,prx,floatfix]{revtex4}

\usepackage{graphicx}

\usepackage{dcolumn}

\usepackage{bm}

\usepackage[usenames,dvipsnames]{xcolor} \usepackage{dsfont}
 \usepackage{amsbsy}

\usepackage{graphicx,bm,times}
 \usepackage{amsmath,amssymb} 
\usepackage{dsfont}
\usepackage{appendix}

\date{\today}

\begin{document}

\title{Dynamical decoupling based quantum sensing: Floquet spectroscopy}
\author{J. ~E.~ Lang} \affiliation{Department of Physics and Astronomy, University College London,
Gower Street, London WC1E 6BT, United Kingdom}

\author{R. ~B. ~Liu} \affiliation{Department of Physics, The Chinese University of Hong Kong, Hong
Kong, China}

\author{T. ~S.~ Monteiro} \affiliation{Department of Physics and Astronomy, University College
London, Gower Street, London WC1E 6BT, United Kingdom}

\begin{abstract} Sensing the internal dynamics of individual nuclear spins or clusters of
nuclear spins has recently become possible by observing the coherence decay of a nearby electronic
spin: the weak magnetic noise is amplified by a periodic, multi-pulse decoupling sequence. However,
it remains challenging to robustly infer underlying atomic-scale structure from decoherence traces
in all but the simplest cases. We introduce Floquet spectroscopy as a versatile paradigm for analysis
of these experiments, and argue it offers a number of general advantages. In particular, this technique generalises to more complex situations, offering
physical insight in regimes of  many-body dynamics, strong coupling and pulses of finite duration. As
there is no requirement for resonant driving, the proposed spectroscopic approach permits physical interpretation of striking, but
overlooked, coherence decay features in terms of the form of the avoided crossings of the underlying
quasienergy eigenspectrum. This is exemplified by a set of ``diamond'' shaped features arising for transverse-field
scans in the case of single-spin sensing by NV-centers in diamond. We investigate also applications
for donors in silicon showing that the resulting tunable interaction strengths offer highly
promising future sensors. 
\end{abstract}

\maketitle

\section{Introduction} 

There is enormous interest in the rapidly advancing field of detection and imaging
at the single spin level \cite{Jelezko2004,Childress2006, Maze2008a,Bala2008},
mainly with NV colour centers but also other defects in diamond \cite{sivac1,sivac2}
not only as a source of versatile qubits for quantum information \cite{NV3, Neumann2010,NV4,Taminiau} and entanglement generation \cite{Bernien2013}, but principally
because they underpin a new generation of quantum sensors, for magnetometry and atomic scale
characterisation of the environment
\cite{Cai,Muller14,Zhao2011,Zhao2012a,Renbao2014,Degen2015,Degen2014}. 
In the widely-studied case of dynamical decoupling quantum sensing, a sequence of pulses 
modulates the coherent evolution of the sensor and in some cases, sharp ``dips'' in coherence allow one
to detect, and infer useful characteristics of, nearby single spins or small spin-clusters. Complexities in the environment 
being studied mean that the single isolated sharp dip is found in a restricted subset of the data and motivates
the development of more general or alternative methods of analysis. 
In particular, many decoupling sequences are temporally periodic.

\begin{figure*}[ht!] 
\includegraphics[width=2.5in]{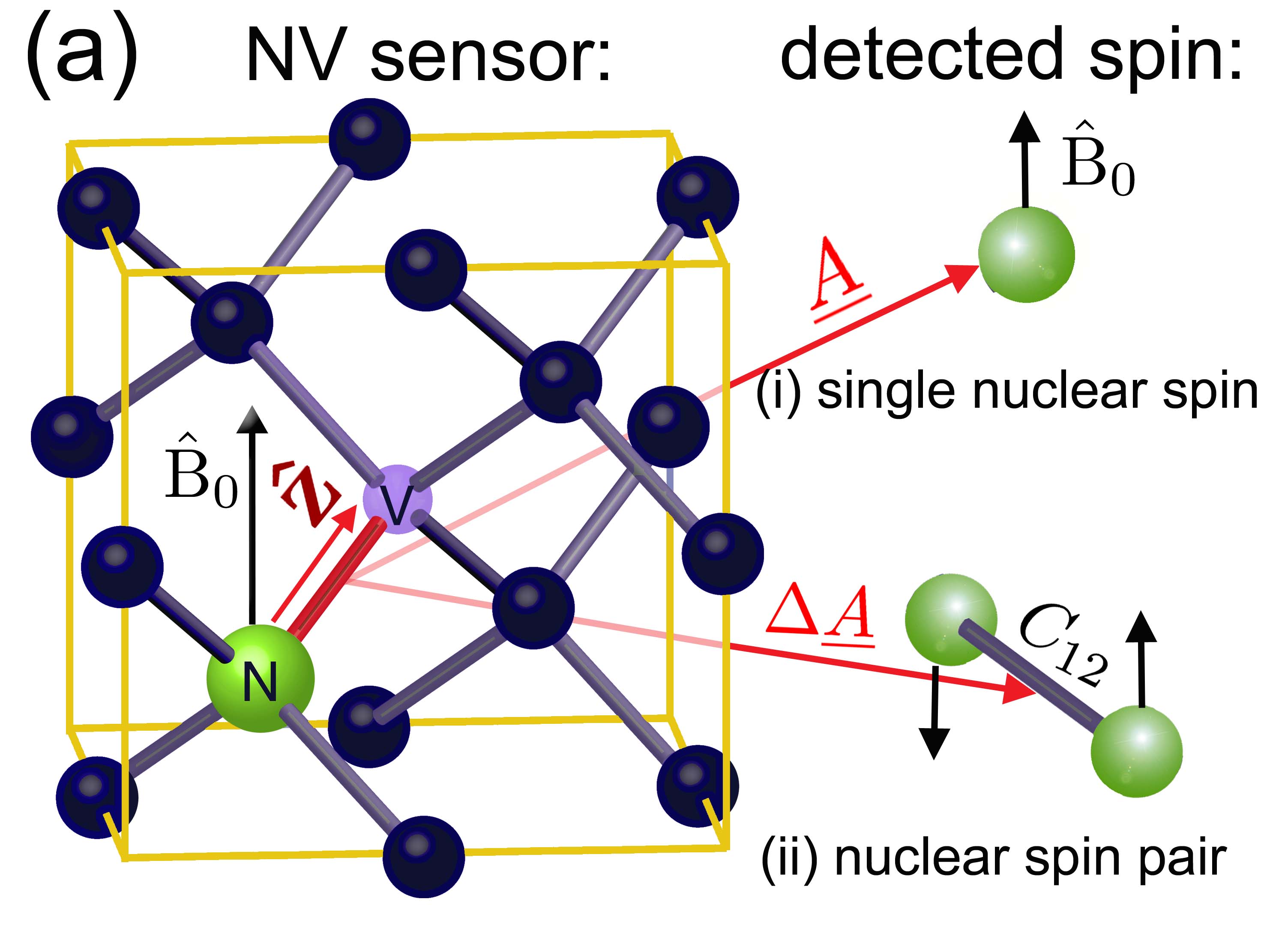}
\hspace{5mm} 
\includegraphics[width=2.in]{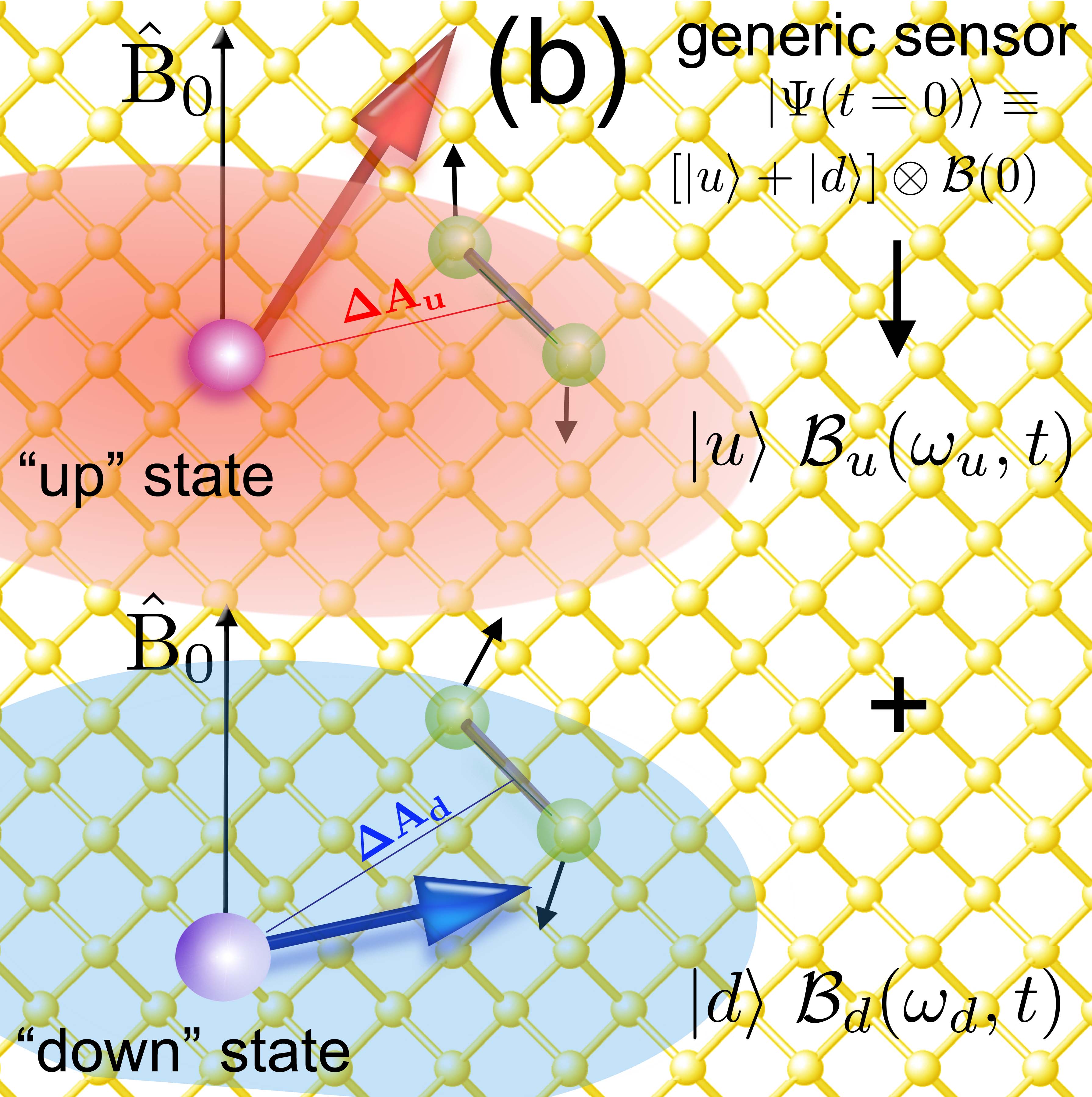} \hspace{5mm}
\includegraphics[width=2.in]{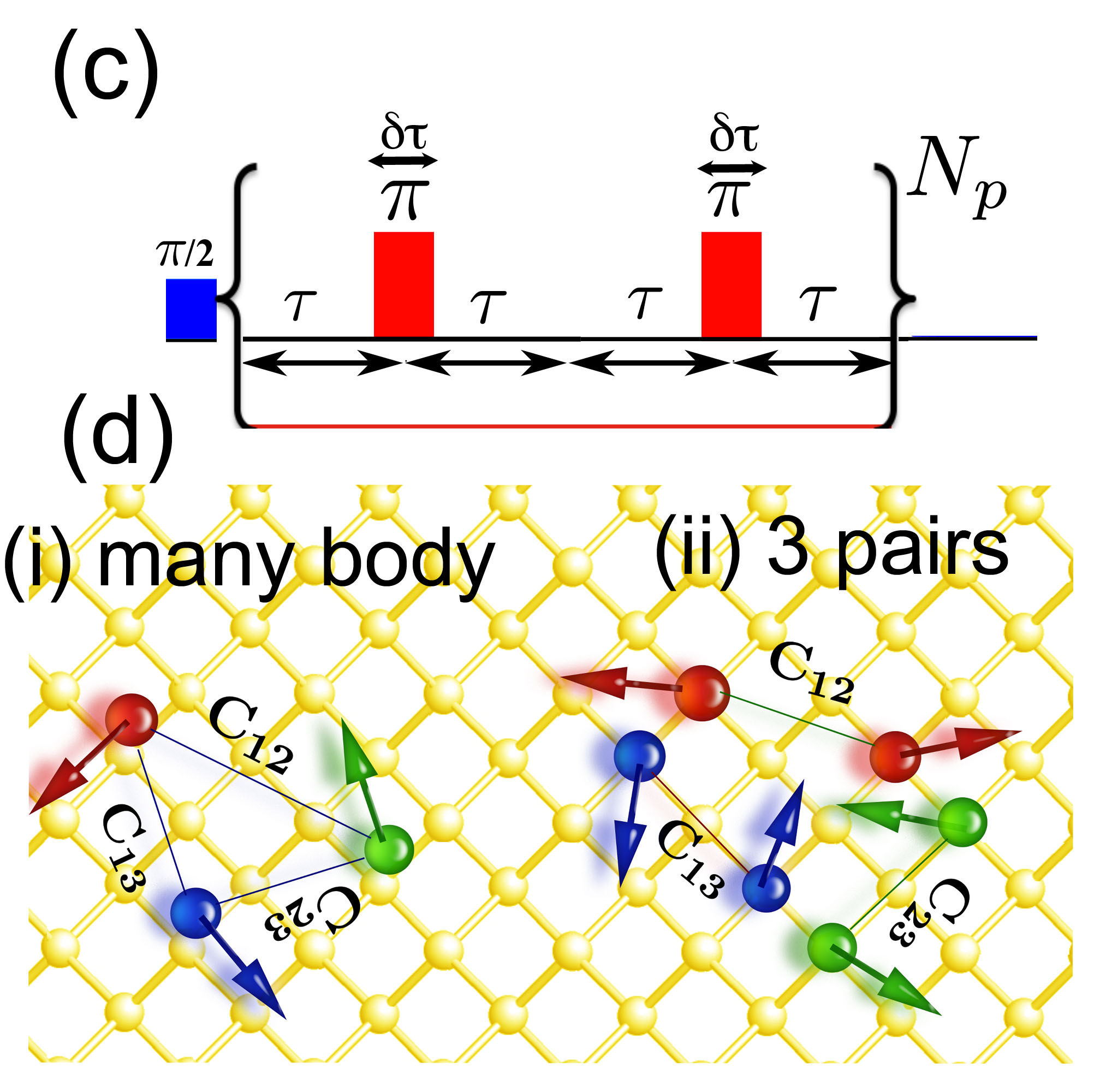} 
\caption{ {\bf (a)} Current experiments have employed the $S=1$
electronic spins of NV-centers to successfuly detect (i) single nuclear spins \cite{Zhao2012a,Kolkowitz2012,Tam2012} (ii) the internal
dynamics of nuclear spin pairs \cite{Renbao2014} as well to characterise on the atomic scale, by
estimating parameters such as electronic-nuclear dipolar couplings $A$ and inter-nuclear dipolar
couplings $C_{12}$. {\bf (b)} Generic sensor detecting a pair of nuclear spins: the electronic spin state
is in a superposition of ``up'' $|u\rangle$ and ``down'' $|d\rangle$ states. The nuclear dynamics
and its characteristic frequencies $\omega_{u,d}$ depend on the electronic state. In turn, the
electronic coherence is sensitive to the resulting weak ac noise from the nuclei. This may be
amplified by dynamical decoupling control such as CPMG, leading to observed ``dips'' in coherence.
These are at well-defined frequencies in typical weak coupling regimes where the nuclear dynamics is
not too different in the $u,d$ subspaces. However, strong coupling regimes do not necessarily yield
sharp dips. {\bf (c)} Additional complexities occur for pulses of finite durations {\bf (d)} It is
is also challenging to differentiate between (i) independent pairs of spins and (ii) many-body
effects from an equivalent interacting cluster. Floquet theory is not restricted to single spins or
spin pairs and can be applied also to analysis of larger, correlated spin clusters, strong coupling
and off-resonant driving.} 
\label{Fig1}
 \end{figure*}

In this case, Floquet's theorem provides a canonical form for the solution of periodically driven systems and has
found wide applicability in various branches of quantum physics since 1965 \cite{Shirley1965},
especially in light-matter interactions with continuous driving and multi-photon atomic physics.
Floquet's theorem can be applied to any
periodic quantum Hamiltonian for which ${\hat H}(t+\tau_{tot})={\hat H}(t)$. In practical implementations,
instead of analysing the eigenstates of the static Hamiltonian, which are appropriate only in the
perturbative limit of weak driving, one employs instead the eigenspectrum
 of the one period time-evolution propagator. Floquet theory is employed in analysis 
 also of Nuclear Magnetic Resonance (NMR) spectra and related \cite{FloquetNMR}, where applications are essentially limited
to  resonant and continuous driving. 

But  this approach has not been 
 considered for analysis of coherence behaviors in this new generation of multipulse spin sensing experiments.
We argue that the Floquet spectroscopy method proposed is better adapted
to regimes of strong quantum entanglement between the sensor and detected spin systems than signal processing methods
applied to classical ac signals; or geometric approaches based on two-state systems.
 In this work, we find that Floquet theory can augment current methods of analysis in several ways:

{\bf (1) } Floquet theory is equally applicable to off-resonant as to resonant driving.
Understanding of spin-sensing data is often cast in signal processing terms: the multi-pulse
sequence imposes a filter function which selects an ac signal with a reasonably well-defined
characteristic frequency $\omega_{ac}/2\pi$ which may, in turn, be used to infer interatomic
coupling parameters. When $\omega_{ac}$ is resonant with the pulse interval $\tau$ (see Fig.~\ref{Fig1}(c)), so when $\omega_{ac}/\pi = 1/(4\tau)$, a
narrow dip in coherence is observed. Here we show that away from such resonances, or even for broadened
resonances, Floquet theory can shed insight on other striking features which are not
narrow dips but  can, nevertheless, yield rich information for atomic scale characterisation.
The key reason is that we show the widths and shapes of these features may be  understood in terms of avoided crossings of an underlying Floquet spectrum.

{\bf (2)} Current experiments operate primarily in regimes of weak-coupling. The pulses involve
consecutive switching between two electronic states $u,d$; the associated characteristic frequencies
of the detected spins $\omega_u,\omega_d$ can vary significantly, since, for stronger coupling,
there is significant back-action and entanglement between the sensor and detected spins. For weak
coupling, $\omega_u \simeq \omega_d$ and in addition average Hamiltonian theory models apply, predicting
typically $\omega_{ac} \approx \frac{1}{2}(\omega_u + \omega_d)$. Floquet theory remains valid regardless of
coupling strength; we examine regimes of failure, obtain alternative forms for $\omega_{ac}$ and
show that the avoided crossings shed insight in these regimes. It remains also valid even if there
is non-trivial evolution due to finite duration of the pulses, a problem only recently identified
\cite{Degen2015}.

{\bf (3)} For detection of two-state systems (spins or spin pairs which reduce to an effective
pseudospin) geometric methods \cite{Kolkowitz2012,Renbao2014} are widely used to interpret data and yield analytical
expressions for the coherence decays. Although here the Floquet method already sheds additional physical insight,
its full value is that it is universally valid even for higher dimensional state-spaces so would facilitate studies of e.g. multi-spin
clusters.

The key-features of dynamical-decoupling based quantum sensing,
using a multipulse periodic sequence, are illustrated in Fig.\ref{Fig1}. 
A $\pi/2$ pulse prepares the sensor system in a superposition state
 $\psi(t=0)=\frac{1}{\sqrt{2}} (|u\rangle+|d\rangle)\otimes {\mathcal{B}}(0)$, where ${\mathcal{B}}(0)$ is the detected spin-cluster at initial time. In turn, 
interaction with the sensor means that the spin-cluster becomes entangled with the sensor
$\psi(t) \simeq \frac{1}{\sqrt{2}} (|u\rangle{\mathcal{B}}(\omega_u,t)+|d\rangle{\mathcal{B}}(\omega_d,t))$,
where $ {\mathcal{B}}(\omega_{u,d} , t)=({\hat T}_{u,d})^{N_p} {\mathcal{B}}(0)$, 
for a pulse sequence (with propagator ${\hat T}_{u,d}$)  which is repeated $N_p$ times. The detected spin dynamics 
is associated with a characteristic frequency which is state-dependent. The temporal coherence ${\mathcal L}(t)= \langle S^+ \rangle$, is  given by ${\mathcal L}(t)= \langle {\mathcal{B}}(\omega_u,t) | {\mathcal{B}}(\omega_d,t) \rangle$  to within a normalisation factor; averaged over bath states, it simulates the experimentally measured signal.

 \section{Floquet Theory}
\begin{figure*}[htb!]
\centering
\includegraphics[width=2.5in]{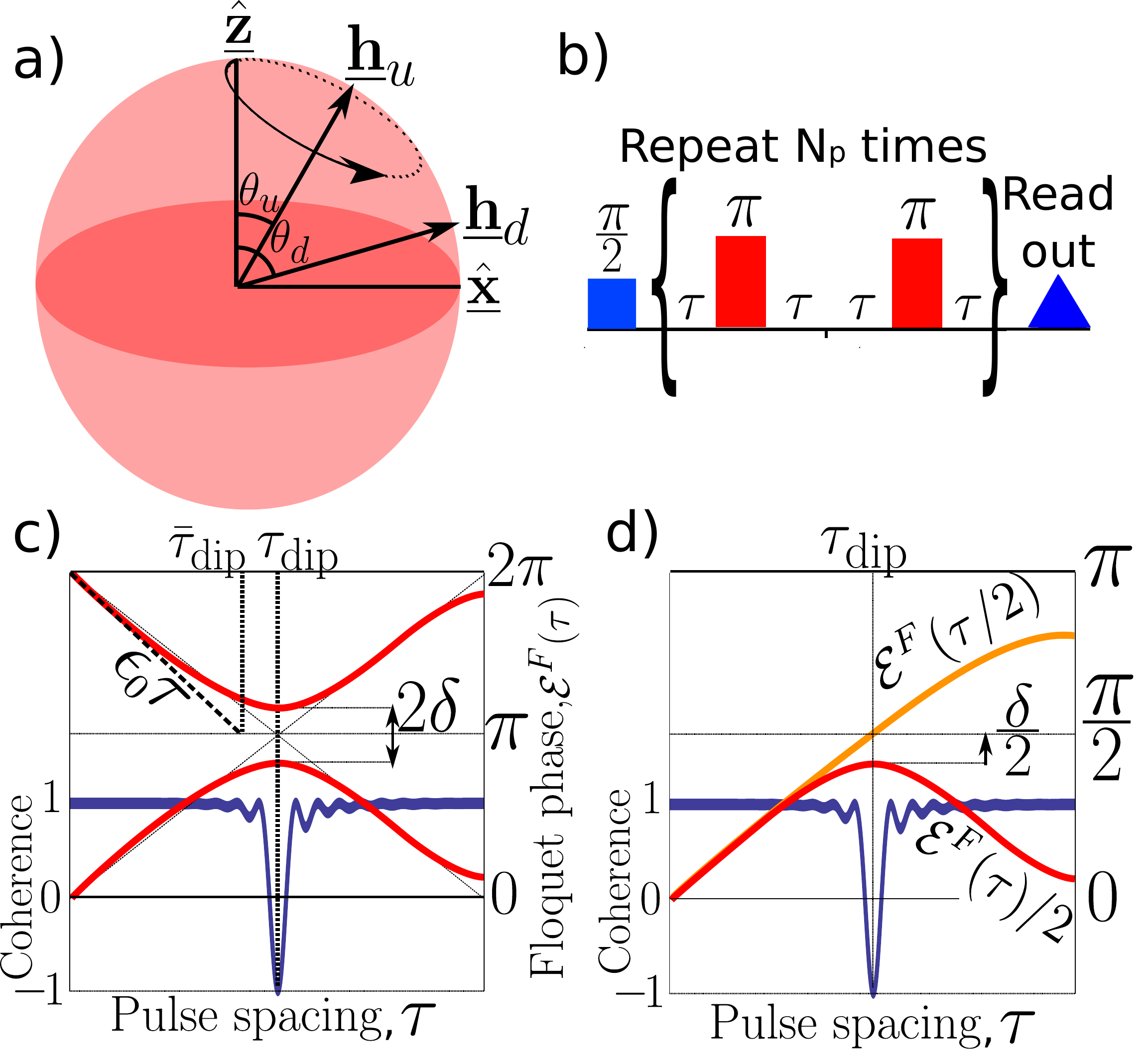}
\hspace{5mm}
\includegraphics[width=4.0in]{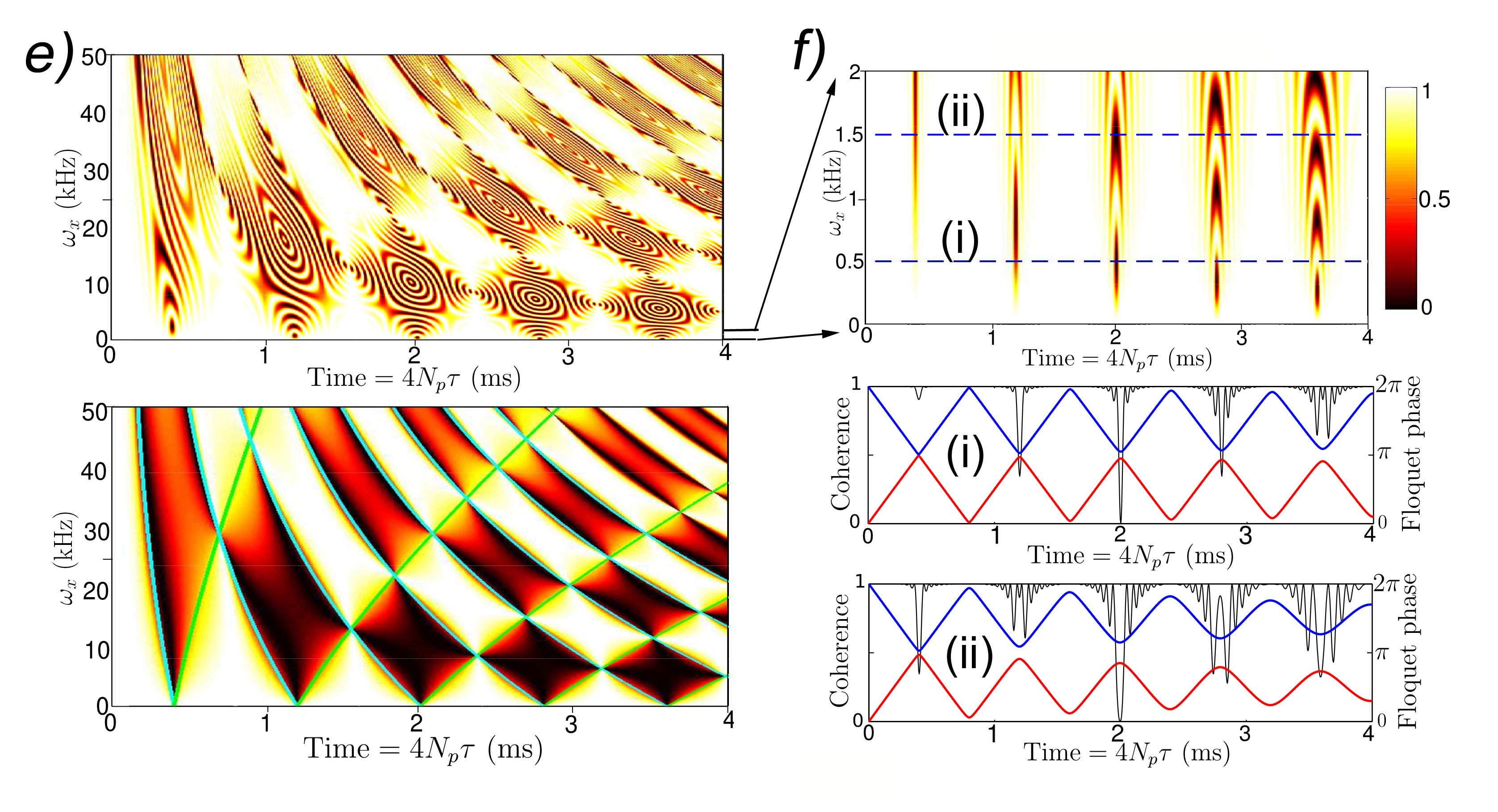}
\caption{ {\bf (a)} usual {\bf Geometric} approach:  under CPMG-N control ({\bf (b)})
the detected spins represent two-state systems which precess about effective magnetic field,  depending on the 
``up''  $|u\rangle$ and ``down''  $|d\rangle$ states of the probe spin. The coherence dips are understood 
by following the precessions and relative angles between these spins, with increasing $N$.
 {\bf (c)} {\bf Spectroscopic} picture. The dips in coherence occur at avoided crossings of the Floquet 
eigenstates. Both the position and contrast of the decoherence dip is related to the curvature of the crossing.
This is characterised by the splitting between the states $2\delta$ and the deviation from the early $\tau \to 0$
 linear evolution. The early time evolution (the $\epsilon_0$ quasienergy) gives  the
the dip position $\overline{\tau}_{dip}$ for 
 average Hamiltonian theory; the coherence minimum is given by $\mathcal{E}(\frac{\tau_{dip}}{2})= \frac{\pi}{2}$.
{\bf (d)} The dip contrast depends on the degree of curvature of the crossing, characterised by
the level-repulsion strength,
$\delta=2\left(\frac{\mathcal{E}(\tau_{dip})}{2} -\mathcal{E}(\frac{\tau_{dip}}{2})\right)$.
 {\bf(e)} NV-center decoherence ``diamonds''. While typical experimental studies
scan along parallel field  ($\omega_z$) component (thus remaining in weak-coupling, single-dip regime),
 scanning the {\em transverse} magnetic field ($\omega_x$) would produce diamond pattern of high decoherence  regions,
  as avoided crossings widen (and even overlap)  then narrow (here $\omega_z = 0$ and $A_{||} = 50$kHz). Upper panel shows full oscillating coherence function, for $N_p=10$
 pulse pairs, lower panel shows coherence envelopes, filled as $N_p\to \infty$. Here $\omega_z=0$.
 Boundaries of the diamonds trace out (green) $\tau=\pi/2(\omega_d+\omega_u)$ 
and (cyan) $\tau=\pi/2(\omega_d-\omega_u)$  (see below). {\bf(f)})
expanded version of low field region showing shape of avoided crossings versus coherence traces corresponding to two cuts (i),(ii), indicated  in the upper panel. }
 \label{Fig2}
\end{figure*}

Floquet's theorem is generally applicable to periodically-forced systems, classical or quantum,
but it allows one specifically to write solutions to the Schr\"odinger equation 
in terms  quasi-energy states (QES) $|\psi_l (t)\rangle= \exp{(-i{\epsilon}_l t)}   |\Phi_{l}\rangle$
where ${\epsilon}_l$ is the quasi energy,
$|\Phi_l(t) \rangle=|\Phi_l(t+\tau_\textrm{tot})\rangle$, $\tau_{tot}$ is the period and
 $l=1,..,D$ ($D$ is the dimension).  
However, for problems (such as our present study)  where we require
only ``stroboscopic'' knowledge of our system (i.e. read-out once every period $\tau_\textrm{tot}$) , the solution is even simpler.
We can obtain Floquet phases/modes simply as the eigenvalues/eigenstates of the one-period unitary
evolution operator ${\hat T}(\tau_\textrm{tot},0)$. The Floquet modes $ |\Phi_l\rangle$, obey the 
eigenvalue equation:
\begin{equation}
{\hat T}(\tau_\textrm{tot})|\Phi_l\rangle = \lambda_l|\Phi_l\rangle \equiv \exp{(-i{\mathcal E}_l)}   |\Phi_l\rangle
\label{Floquet}
\end{equation}
 where now ${\mathcal E}_l$ is the  eigenphase (the Floquet phase) and  $\epsilon_l={\mathcal E}_l/\tau_\textrm{tot}$
is the quasienergy. For sensing, we can obtain Floquet phases/modes simply as the eigenvalues/eigenstates of
 ${\hat T}_{u,d}$, the basic 
periodic sequence; for example, for the CPMG sequence in Fig.\ref{Fig2},  $\tau_\textrm{tot}=4\tau$.
In that  instance ${\hat T}_i | \Phi_{il}\rangle=e^{-i {\mathcal E}_l^{(i)}}|\Phi_{il}\rangle$ where $i=u,d$ denotes the state of the sensor spin. The eigenphases for  $N_p=1$ fully determine the time-evolution of the system
since if the pulse sequence is repeated $N_p$ times,
we just scale the eigenphases, so for longer time-propagation:
\begin{equation}
\left(\hat{T_i} \right)^{N_p}| \Phi_{il}\rangle=e^{-i N_p{\mathcal E}_l^{(i)}} |\Phi_{il}\rangle.
\end{equation}

In the present work, we show that these Floquet eigenphases and eigenstates have particular important properties:\\

\begin{figure}[ht!]
\includegraphics[width=3.5in]{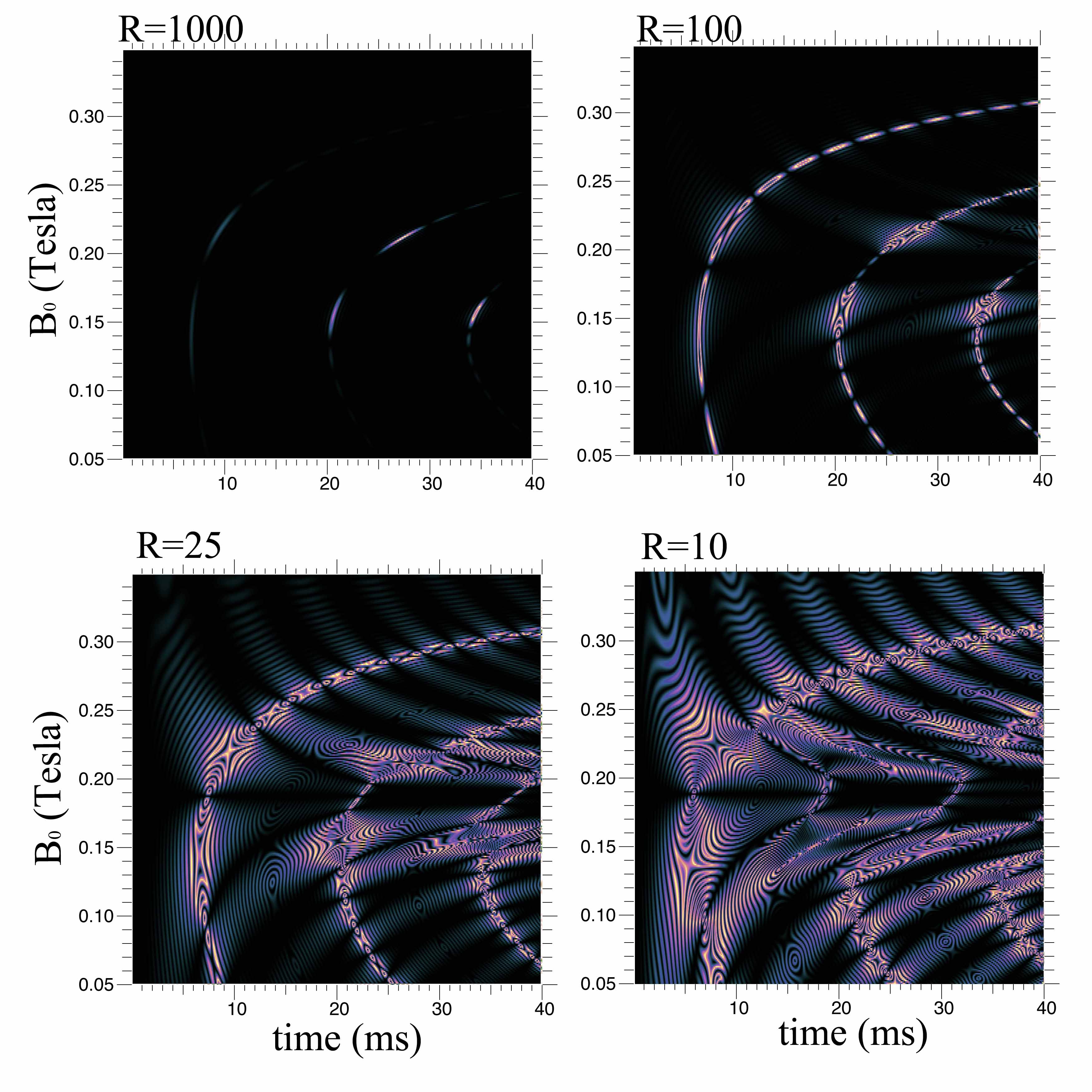}
\caption{Coherence decay behavior for an electron spin detecting a flip-flopping pair of nuclear spins,
for a donor in silicon system (see \cite{SuppInfo}) with tunable interactions.
$\mathcal{L}(B_0, t)$ exhibits a rich structure in the two-dimensional  $ \tau, B_0$ plane 
which is  not evident in the  normal traces at constant $B_0$. Decoherence map is shown for different $R=\Delta{}A/C_{12}$, $(\Delta{}A = A_1-A_2)$.
Large $R$  corresponds to weaker dipolar coupling $C_{12}$ and the maps trace the locus of
a set of isolated sharp dips in coherence.
For smaller $R$, there are no longer single dips; nevertheless the envelopes (given by  $F(\tau)$) are well-defined
and track the behavior of the underlying Floquet avoided crossings.
 The background oscillatory structure depends on $N_p$, the envelopes do not. Time $t \equiv 4N_p\tau$;
 (colour scale  linear, with black $\equiv 1$, yellow $ < 0.5$). Similar behavior is obtained for
several transitions of Si:Bi and
other donors, but specific parameters are for
   $12 \to 9$  ESR transition of Si:Bi and  $2N_p=40$.}
\label{Fig3}
\end{figure}

{(\bf a) } The eigenvalues  are the same for the upper and lower
 states, i.e.  $e^{i{\mathcal E}_l^{(u)}}=e^{i{\mathcal E}_l^{(d)}} \equiv e^{i{\mathcal E}_l}$.
 This holds even for pulses of finite duration,
in typical cases. In other words, the evolution of the ${\hat T}_u^{(N_p)} {\mathcal{B}}(0)$ and ${\hat T}_d^{(N_p)} {\mathcal{B}}(0)$
are characterised by the same set of effective frequencies
$\epsilon_l={\mathcal E}_l/\tau_\textrm{tot}$, in contrast to typical static, 
geometric approaches where two distinct sets of  frequencies $\omega_{ul}$ and $\omega_{dl}$ are involved.\\
 {(\bf b) }  The eigenvectors do not, in general, coincide
but we show that (to within a phase term),  the eigenvectors are related to each other by a half-period evolution
 e.g. $ {\hat T}_u(\tau_{tot}/2) | \Phi_{ul}\rangle \propto | \Phi_{dl}\rangle$.\\ 
{(\bf c) } Minima in coherence (of prime importance for sensing, whether sharp dips or not)
 occur at avoided crossings of Floquet eigenstates, where
 $e^{i{\mathcal E}_l} \simeq e^{i{\mathcal E}_k}$. 
Once the Floquet phases and modes are obtained, one can obtain the general form of the decoherence for
arbitrary times which, averaged over bath states, yields:
 \begin{eqnarray}
 \langle\mathcal{L}(t=N_p4\tau)\rangle = \frac{1}{D}\sum^D_{l,l'} e^{-i N_p ({\mathcal E}_l-{\mathcal E}_{l'}) } | \langle \Phi_{dl'} | \Phi_{ul} \rangle|^2 
\label{Fdeco}
 \end{eqnarray}

Derivations of  {\bf{(a)-(c)}}  are given in the Appendix. Although properties (a)-(c) are quite generic, physical insight on the Floquet picture is more easily gained from two-state systems, where 
direct comparison with usual geometric methods \cite{Zhao2012,Kolkowitz2012,Ajoy2015}
is also possible. For the two-state case, 
eigenvalues must be conjugate pairs $\lambda_\pm= e^{\pm i{\mathcal E}}$. Level
 crossings  occur when $\lambda_+ \simeq \lambda_-$ hence the crossings must occur at ${\mathcal E} \simeq 0, \pi, 2\pi$. The generic properties of states at avoided crossings  (see Appendix) then imply coherence dips occur at ${\mathcal E} \simeq \pi$.
We now first investigate the Floquet dynamics for these two-state single-spin or single pseudospin models.

\begin{figure}[ht!]
\centering
\includegraphics[width=2.9in]{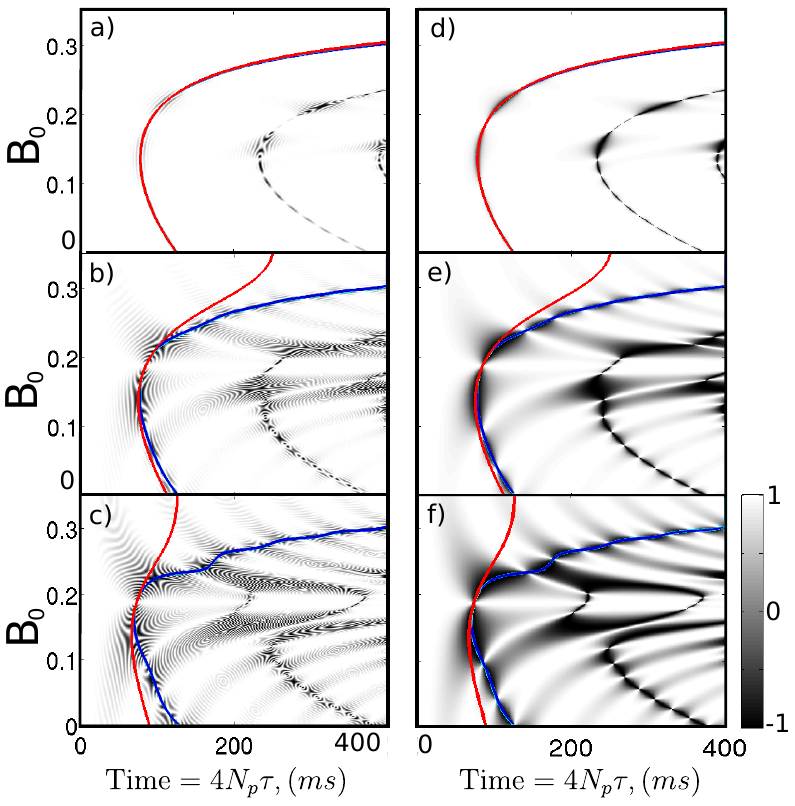}
\caption{Comparisons between the dip positions obtained with Eq.\ref{Lspec} 
 (blue line) and average
Hamiltonian theory Eq.\ref{avdip} (red line) for the full coherence function 
(left panels) as well as its envelope (right). Within the field sweep there are global
 weak-coupling points (eg $B_0 \approx 0.19$ T) where 
there is weak coupling regardless of the cluster properties and where the decoherence envelope collapses into 
a single sharp dip, a useful feature if high resolution is required:
 here there is always good agreement with average Hamiltonian theory.
These points correspond to so called optimal working points \cite{Mohammady2010,
Balian2014} of silicon donors.
Hence, the advantage of such systems as future spin sensors,
 in addition to their very long $\sim 1$ s coherence times,
is that a magnetic field sweep could tune the dynamics from the
weak to strong coupling regimes.
(a) and (d) R=100 (b) and (e)  R=20 (c)  and (f) R=10.}
 \label{Fig4}
\end{figure}

\subsection{Single spin or spin pair detection} 
\label{onespin}
Both pair flip-flop dynamics as well as single spin-dynamics  (in systems like NV centres where a crystal field leads to non trivial one-spin
dynamics) can be approximated by a two-state Hamiltonian. We term this the pseudospin model, noting that for  single-spin detection  there is a genuine spin, while for pair-dynamics  \cite{Zhao2011,Renbao2014} it is a pseudospin. It has led to a successful, widely used geometric model (see Fig.\ref{Fig2}(a)  and \cite{SuppInfo}) where
the  evolution of the pseudospin is conditional on the state $i=u,d$ of the probe and corresponds
to precession about an effective magnetic field:
$H_i=\frac{1}{2} {\textbf{h}}^{i} \cdot {\pmb{\sigma}}=\frac{1}{2}(X\sigma_x+ Z_i\sigma_z)$
where $\sigma_x,\sigma_z$ are Pauli matrices in the usual spin basis; in the  pseudospin case of course,  we have
$|\uparrow\downarrow\rangle \rightarrow |\uparrow\rangle$ and $|\downarrow\uparrow\rangle\rightarrow|\downarrow\rangle$). 
The $X,Z_i$ depend on the physical system (see \cite{SuppInfo} for details);
but for NV centers   $\textbf{h}^{u} \simeq (\omega_x,0,  A_{\parallel}+\omega_z)$ while $\textbf{h}^{d} \simeq (\omega_x, 0,  \omega_z)$ 
where $\frac{\mu_0 {\bf B}_0}{\hbar}=(\omega_x,0,\omega_z)$ is the external magnetic field,
and $A_{\parallel}$ the parallel component of the hyperfine interaction.
For spin pair-sensing, on the other hand,  $\textbf{h}^{i} = \frac{1}{2}(C_{12},0, \Delta A_i)$
where $\Delta A_i=2(A_1-A_2) \langle i | {\hat S}_z | i \rangle$  represents the energy detuning between the nuclear spins in the pair and $\hat{S}$ represents the operator for the sensor spin.
The eigenvalues of $H_i$  are
$\omega_{u,d}= \pm \frac{1}{2} \sqrt{X^2+Z_{u,d}^2}$
and the orientation of the effective field is $\theta_i = \arctan(X/Z_i)$.
For two state systems,  we obtain:

\begin{figure*}[ht!]
\centering
\includegraphics[width=8.in]{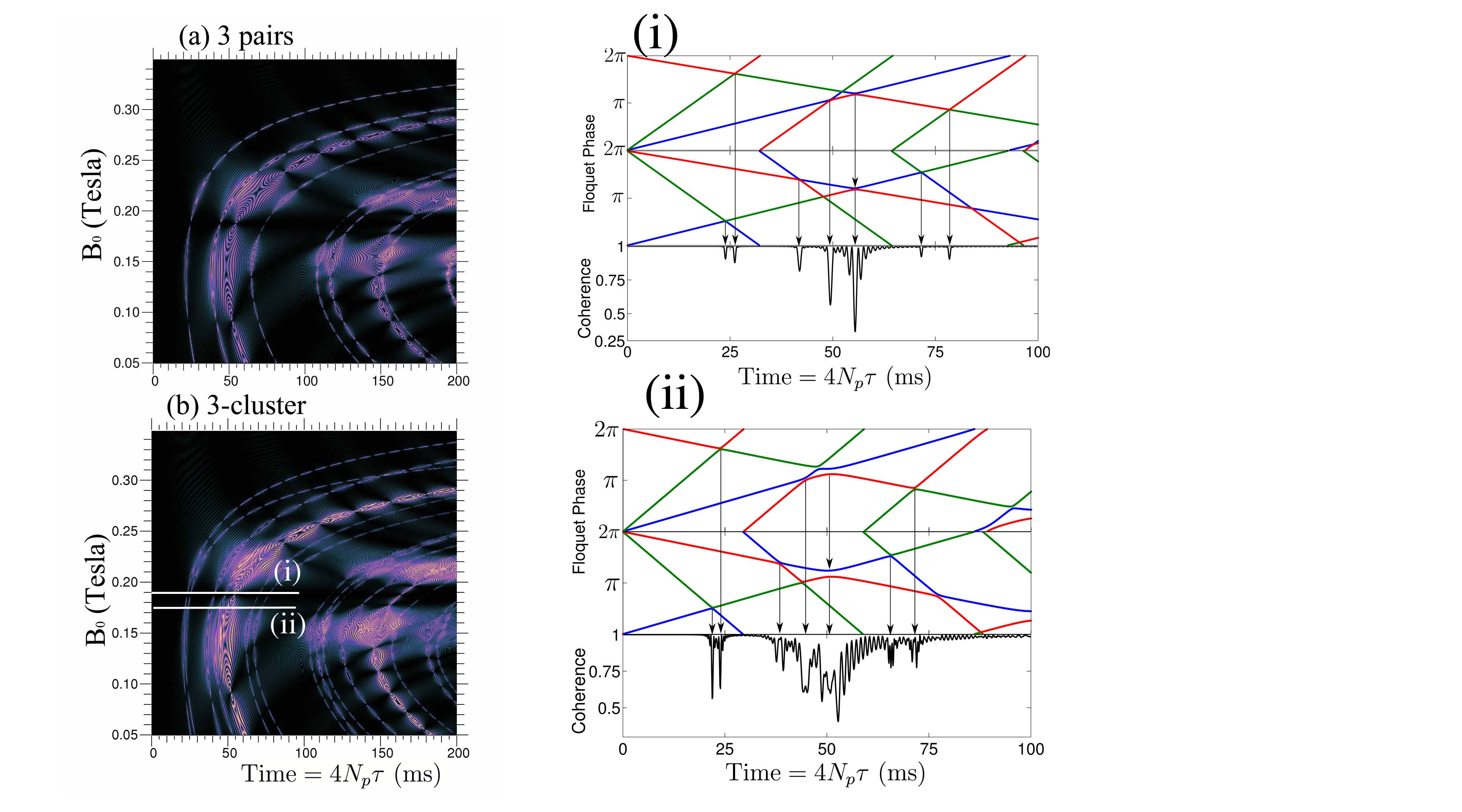}
\caption{Fingerprinting multiple environmental spin cluster-pairs via their decoherence``bar-codes''
 illustrates the effect of 3-body correlations. The figure shows the coherence as a 
function of magnetic field $B_0$ and pulse interval $\tau$, calculated with a full numerical propagation
under the total Hamiltonian for $N_p=100$. {\bf{(a)}} denotes three independent pairs while {\bf{(b)}}shows three interacting spins,
with otherwise equivalent dipolar couplings and intrabath interactions as illustrated in Fig.\ref{Fig1}(b).
One evident difference (and signature of a cluster of three spins) are the doublets due to the two separate subspaces of the three interacting spins.
The splittings are directly related to the interactions.
For the 3-cluster, in fact there is a secular contribution from interactions between spins, greatly amplifying their contribution.
The two right hand panels (i) and (ii) show single traces corresponding to the cuts in {\bf{(b)}} as well as the six
corresponding eigenphases:  in case (i) in a weak coupling regime the dips are narrow and the eigenphases behave
like three independent pairs; the eigenvalues correspond to conjugate pairs (with blue, red and green lines denoting 
the three pairs). In case (ii) there is stronger coupling, the avoided crossings
of the corresponding eigenphases are broader giving raise to the features shown in the coherence maps. }
 \label{Fig5}
\end{figure*}

\begin{align}  
 {\langle\mathcal L}(\tau)\rangle  =  1 - 2 \left[\frac{\cos^2 { [\mathcal{E}(\tau)/2]} -\cos^2 {[\mathcal{E}\left(\tau/2\right)]}}
{ \cos^2 { [\mathcal{E}(\tau)/2] }} \right ] \sin^2 \left[N_p\mathcal{E}(\tau)\right] 
  \label{Lspec}
\end{align}
This is a key result of the present work as it means one can give the full coherence function using only the Floquet phases. To calculate the Floquet eigenphase $\mathcal{E}(\tau) $, as well as its half-period value 
 $\mathcal{E}\left(\tau/2\right)$ in Eq.\ref{Lspec},
one may use
$\cos({\mathcal{E}(s)})  =  \cos(2\omega_u s)  \cos(2\omega_d s)-\sin(2\omega_u s)  \sin(2\omega_d s)  \cos(\theta_u-\theta_d)$ with $s=\tau$ or $s=\tau/2$. Thus the coherence takes the form
 $ {\mathcal L}(\tau)  =  1 - 2F(\tau) \sin^2 \left[N_p\mathcal{E}(\tau)\right]$, which is the product of a smooth envelope
$F(\tau)$, independent of $N_p$, superimposed on a fast oscillating function $\sin^2 \left[N_p\mathcal{E}(\tau)\right]$, dependent on $N_p$.

Full comparison with geometric methods are in \cite{SuppInfo} where we argue  it is the
 condition 
\begin{equation}
\mathcal{E}(\tau_{dip}/2)= \pi/2
\end{equation}
 which best specifies the dip positions. The {\em depth} of the dip  is related to the 
eigenvalue splitting parameter
$\delta= \pi- \mathcal{E}(\tau_{dip}) $; at the dip 
\begin{equation}
 \mathcal{L}(t=4N_p\tau_{dip}) =  1 - 2  \sin^2(N_p \delta).
\end{equation}
Hence, if $ \mathcal{E}=\pi$, there is no dip, so a true level crossing provides no signal.
For  $N_p\delta \gtrsim \pi/2$ the width and shape of the dip becomes largely independent of $N_p$
and is fully determined by the Floquet anti-crossing and  envelope function, since the $\sin^2(N_p \delta)$ prefactor simply superposes fast oscillations on  $F(\tau)$. A narrow avoided-crossing (low splitting, $\delta$ small) gives a single, sharp (but weaker) coherence dip, while a large $\delta$ crossing has a broad envelope.
It is only for low  $N_p\delta \ll \pi/2$ that the dip height is strongly dependent on $N_p$; here the central height increases  as $(N_p\delta)^2$.

{\bf Comparison with Average Hamiltonian models}  (see \cite{SuppInfo} for details).
A frequently used approximation in spin sensing is the average Hamiltonian model whereby the
eigenvalues $\omega_\textrm{av}$  of, $\frac{1}{2}(H_{u}+H_{d})$,  the time-averaged Hamiltonian
provide an estimate of the dip positions and that $T_{dip}= \frac{N_p\pi}{\omega_\textrm{av}}$.
From Fig.\ref{Fig2} (c) and (d) we can equate linear behaviour in our eigenvalues (narrow crossing, linear shape)
both with the occurrence of  a sharp dip as well as validity of the averaged Hamiltonian model. 
 In particular, for small $\tau$,  $\mathcal{E}(\tau) \simeq 4 \epsilon_0 \tau$ corresponds to the averaged Hamiltonian
results. Expanding the $\cos{\mathcal{E}(\tau)}$ from below Eq.\ref{Lspec} , for small $\tau$, we easily obtain
$\epsilon_0= \frac{1}{2}(\omega_u^2+\omega_d^2+ 2 \omega_u \omega_d \cos{(\theta_u-\theta_d)})^{1/2}$ and thus:
\begin{equation}
\overline{\tau}_{dip} = \frac{\pi}{2(\omega_u^2+\omega_d^2+ 2 \omega_u \omega_d \cos{(\theta_u-\theta_d)})^{1/2}}
\label{avdip}
\end{equation}
Expressing quantities in terms of the pseudofield components $X,Z_{u,d}$ we can show that this is equivalent
to the expression $\omega_{\textrm{av}} = \frac{1}{2}((X^2 + (\frac{Z_u+Z_d}{2})^2)^{1/2}$ used in spin-detection
experiments \cite{Zhao2011,Renbao2014} and
 to $\omega_{\textrm{av}} \simeq \frac{1}{2} (\omega_u +\omega_d)$ for $\theta_u \simeq \theta_d$.

\section{Applications}
\subsection{Experimental control of quasienergy crossings}
The above motivates us to investigate possibilities for experimental control 
of the avoided crossings, by varying $\delta$,  even in the simple
one-spin or spin-pair case. In typical sensing with NV-centers, we have $\omega_z \gtrsim \omega_x \gg A$,
thus in Eq.\ref{avdip}, we have $\theta_u \sim \theta_d \ll \pi$, thus 
$\overline{\tau}_{dip} \simeq \frac{\pi}{2(\omega_u+\omega_d)}$. However, setting $\omega_z=0$ and increasing
$\omega_x$ causes the anti- crossings to widen and narrow successively, forming a checkerboard pattern of diamonds. This behavior was illustrated in Fig.\ref{Fig2} (e) and (f). In particular Fig.\ref{Fig2} (e) 
illustrates the usefulness of the 2D map; it is not easy to clearly discern the behavior from an
individual trace (as in Fig.\ref{Fig2} (f)) especially if $N_p$ is not very large.
We note that higher harmonics have larger $\delta$ than lower harmonics at the same parameters.

However, here we consider in addition $S=1/2$ systems as potential sensors. These might include silicon vacancies
but in particular we focus on electron donors in silicon. Although techniques analogous to
optical  read-out and polarization of NV centres are not fully developed, there has been considerable 
progress in single-spin detection \cite{Morello2010,Pla2013,Muhonen2014} . These systems benefit also from extremely long coherence
times (of order seconds) for cryogenically cooled samples. They are also an ideal test-bed
for the theory as one can vary $\theta_u, \theta_d$ over a wide range as magnetic field $B_0$ is swept.
For  donor systems, the surrounding  $^{29}$Si nuclear spin dynamics does not generate 
an ac signal as there is no internal crystal axis, in contrast to the case of NV centers in diamond where the 
surrounding nuclear spins precess around an effective quantisation axis which is no longer only the external magnetic field. 
For donors, single  strongly coupled  $^{29}$Si nuclei have recently been detected via the static shift of the donor frequency \cite{Pla2014}.
However, the interesting coherence dynamics in these systems involve only pairs or larger clusters of spins where the flip-flopping dynamics generates an ac signal \cite{Renbao2006,Balian2014}. 

Formally, the state-conditional dynamics for donors is very similar to that for NV centers:
The dynamics correspond to an effective spin precessing about effective magnetic fields 
  $\textbf{h}^{i} = \frac{1}{2}(C_{12},0, \Delta A_i)$
where $\Delta A_i=(A_1-A_2) P_i$; But in contrast to NV centers
 $P_i(B_0) =2\langle i | {\hat S}_z | i \rangle$,
the polarisation of the state (see \cite{SuppInfo}) varies strongly with the magnetic field \cite{Mohammady2010, Morley2013},
while for NVs, $\langle i | {\hat S}_z | i \rangle=0,\pm 1$ is fixed for the modest fields used
in experiments. 

Fig.\ref{Fig3} shows the field-dependence of the coherence for a variety of coupling strengths.
The behavior may be compared with the NV-centers: in this case, the coherence dips trace out a curved locus.
 like NVs, for stronger $X \equiv C_{12}/2$ component of the pseudofield, the envelopes broaden, but there is a similar
pattern of intermittent broadening and narrowing. There is a striking feature at ($B_0=188$mT in  Fig.\ref{Fig3}
where all decoherence envelopes ``collapse'' to a sharp dip. This is one of a set of special fields (Optimal working points) where  $\theta_u \simeq \theta_d$ and $\omega_u \simeq \omega_d$ and
which have been investigated theoretically and experimentally for their favourable coherence properties
 \cite{Mohammady2010,Wolfowicz2013,Balian2014}. 

 But, in the present work, we find also
that these points correspond also to very narrow Floquet avoided crossings, at which $\delta \to 0$.
 Fig.\ref{Fig4} compares dip position predicted by average Hamiltonian theory Eq.\ref{avdip} with 
the accurate dip condition $\mathcal{E}(\tau_{dip}/2)= \pi/2$. 
In Fig.\ref{Fig4}, $\Delta A = (A_1 - A_2)$,
(which approximately sets the timescale for $R \gg 1$)
was fixed, while $C_{12}$ (the intra-bath dipolar coupling) was varied to obtain different values of $R=\Delta A/C_{12}$.

By means of a detailed theoretical analysis we can show that average Hamiltonian theory is valid if 
(i)$ |\mathbf{h}_u + \mathbf{h}_d| \gg |\mathbf{h}_u - \mathbf{h}_d|$ or if (ii) $ |\mathbf{h}_u - \mathbf{h}_d| \gg |\mathbf{h}_u + \mathbf{h}_d|$.
Condition (i) corresponds to the weak-coupling regimes typical of NV sensing experiments, where $\omega_u \simeq \omega_d$ and 
$\theta_u \simeq \theta_u$; it is also the regime of the Optimal Working Points, where average Hamiltonian theory is always valid.
Regime (ii) is not typical of sensing experiments;  for the spin-$1/2$ system of Fig. \ref{Fig4}, it would correspond to the spins nearly antialigned,
thus $P_u \simeq -P_d$. For Fig. \ref{Fig4} condition (ii) implies $(\Delta A)^2(P_u-P_d)^2 \gg (C_{12})^2+ (\Delta A)^2(P_u+P_d)^2$.
In particular, for  $P_u \approx -P_d$ we obtain the condition $|\Delta A(P_u-P_d)| \gg |C_{12}|$. Noting that $|P_u-P_d |\simeq 0.2-2$,
this means that average Hamiltonian theory is valid at all fields for large $R \gtrsim 100$ as seen in Fig. \ref{Fig4}(a).
However, for increasing intra-bath dipolar coupling $C_{12}$, the theory ceases to be valid away from the small OWP regime
in the center, as seen in Figs. \ref{Fig4} (b) and (c) for smaller $R$.

\subsection{Detection of multi-spin clusters}
In this section we apply the Floquet approach to the system depicted in Fig.\ref{Fig1}(d): we compare 
the decoherence ``fingerprint'' of three independent spin pairs (analogous, formally, to
 the detection of three independent spins
by NMR) with a 3-cluster which, in the absence of many-body interactions would give a 
similar signature.\\
{\em For the 3-cluster}, we take three spins, with hyperfine couplings $A_k\equiv A_1,A_2,A_3$ to the
 sensor spin and with mutual
dipolar interactions $C_{ij}\equiv C_{12},C_{23},C_{31}$. Disregarding interactions,  
the energy cost of the spin flips is $\Delta_{ij}=A_i-A_j$.

\begin{figure}[ht!]
\centering
\includegraphics[width=2.6in]{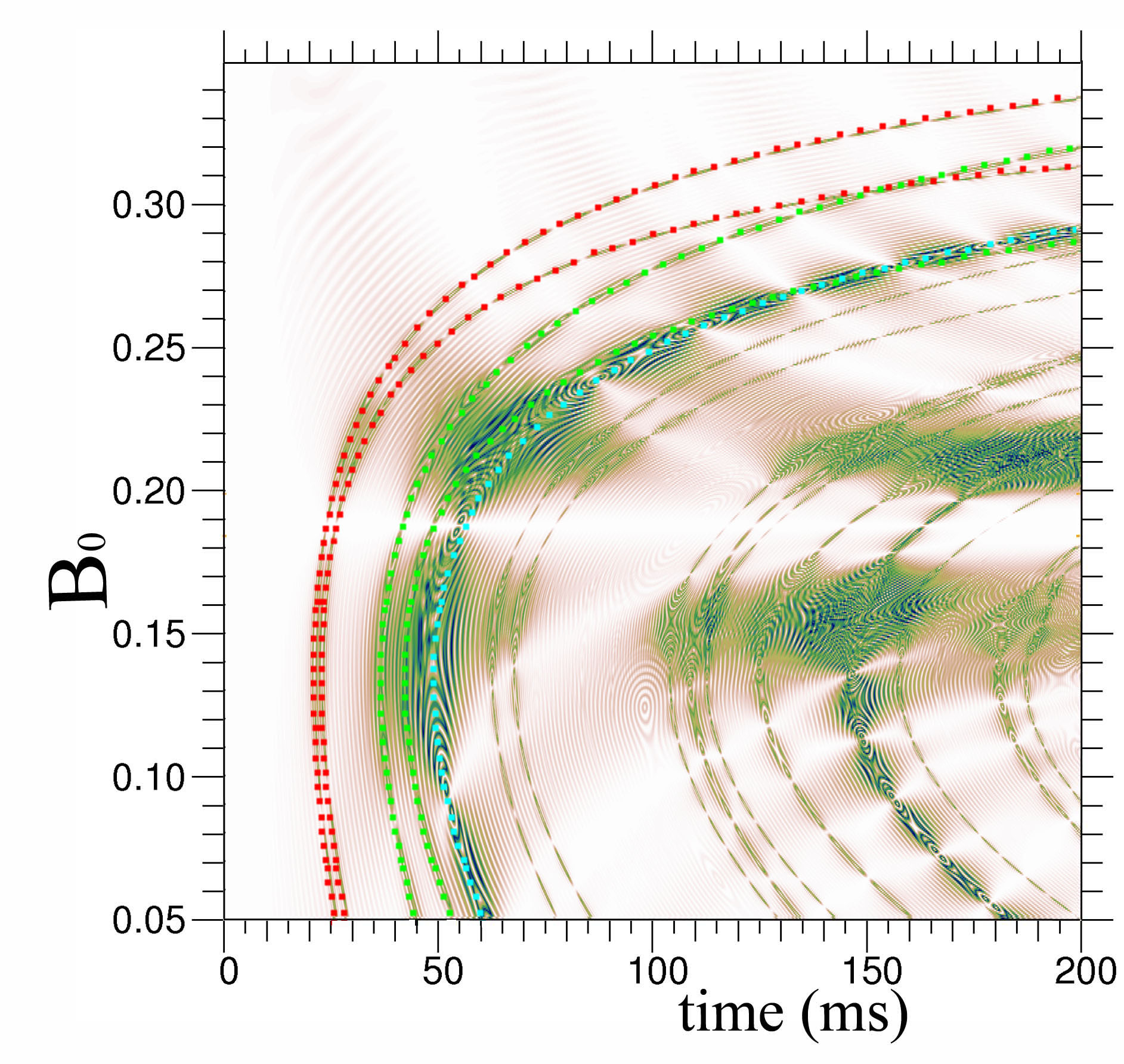}
\caption{ Decoherence for an interacting cluster of three spins (3-cluster) . The coloured lines show comparisons with 
Eq.\ref{eq:secdip2} showing excellent agreement with numerics obtained by diagonalisation of the full
joint sensor-cluster Hamiltonian.}
 \label{Fig6}
\end{figure}

{\em For the independent pairs}, we take three spin pairs, with the  same dipolar interactions $C_{ij}$ 
as the 3-cluster, 
but which are independent of each other. To have similar frequencies as the 3-cluster, 
we must have similar energy cost of a 
all three spin flips; and they must obey the cyclic condition of the 3-cluster
 $\Delta_{12}+\Delta_{23}+\Delta_{31}=0$.
Pair 1 has two spins with interaction $C_{12}$ and a pair of hyperfine couplings $(A_1,A_2)$;
pair 2 has interaction $C_{23}$ and hyperfine couplings $(A_2, A_3)$; pair 3 has $C_{31}$ 
and hyperfine couplings $(A_3, A_1)$. We take $C_{12}=C_{23}=\frac{1.05}{ 2\pi}$~kHz and 
$C_{31}=\frac{2.2}{2\pi}$~kHz, realistic
values for nuclear impurities in the silicon lattice. We take $A_1=\frac{180}{2\pi}$~kHz,
 $A_3=\frac{100}{2\pi}$~kHz and $A_2=0$,
thus our pairs correspond to $R \simeq 100-40$ (as defined in Fig.\ref{Fig3}) so the interactions are sufficiently
 weak to make their
detection challenging but sufficiently strong to, below, illustrate important features.
 The choice of $A_2=0$ does not involve 
much loss of generality. If a state-dependent Hamiltonian
is chosen, the $A_1,A_2,A_3$ values can be shifted by an arbitrary constant without
 perturbing the dynamics.
If the full Hamiltonian is considered, there can be higher order effects such as
 hyperfine mediated corrections
to $C_{ij}$. This correction is very small for our parameters but is tested by full numerics below.

{\bf Solution of total Hamiltonian}
First we set aside all pseudospin approximations and do diagonalisations of the full Hamiltonian
 followed by time-propagation, using the complete 8-state basis of the 3-cluster
as well as the complete basis of the bismuth sensor including the host nuclear spin. Thus, 
unlike Fig.\ref{Fig3}, we do not use the analytical form for the parameter $P_i$; 
it emerges from the numerics. We evaluate the decoherence numerically
rather than using Eq.\ref{Lspec}.  A similar calculation was carried out 
with the three disjoint pairs, then the decoherence
was averaged over the thermal ensemble of nuclear spin states 
(of which there are eight for the 3-cluster).
 Figure \ref{Fig5} shows  maps of the coherence in the $(\tau,B_0)$-space in both cases.

One conclusion to be drawn from comparisons between full numerics and the analytical
 (one-pair) Eq.\ref{Lspec} is that the 
structure in Fig.\ref{Fig3} is surprisingly robust; without bath state averaging,
 full numerics give similar structure to
Fig.\ref{Fig3} (obtained from Eq.\ref{Lspec} for one bath state). 
 
One striking feature of the 3-cluster decoherence map in Fig.\ref{Fig5}(b) is that some
 lines are split into ``doublets'' with very similar
structure. The origin of these is in the average over the bath states;
 examining maps for the individual 8 bath states, we see that while
the $I_z=\pm 3/2$ cluster states $|\uparrow \uparrow\uparrow\rangle$ and 
$|\downarrow \downarrow\downarrow\rangle$ make no
appreciable contribution, the doublets arise from the separate $I^z=\pm 1/2$ subspaces,
 which do not mix.
In other words, the  $|\uparrow \uparrow\downarrow\rangle$,  $|\uparrow \downarrow\uparrow\rangle$
 and $|\downarrow \uparrow\uparrow\rangle$  states with total quantum number $I^z=+1/2$
interact only weakly with the equivalent $I^z=-1/2$ subspace, but each provides a 
locus of dips with a slightly different shift.
In contrast to the spin pairs, in the case of the 3-cluster, the secular 
Ising ($C_{jk} {\hat I}_{zj}{\hat I}_{zk}$) components  yield a non-trivial dynamical effect.

Fig.\ref{Fig5}(i) and (ii) also shows a cut for two field values and compares with the behavior of the Floquet eigenstates. We can see that near the ``weak-coupling'' regime of optimal working points (i),
the dips are sharp and narrow as are  the avoided crossings; in contrast, away from the OWP
point,  avoided crossings are broader and even overlap (ii); the level splitting is much larger. We can estimate  the point where two eigenvalues will collide, and hence $\tau_{dip}$  by exploiting the fact that in either case, the average Hamiltonian theory value  is not too 
far from the accurate value ${\overline \tau}_{dip} \simeq \tau_{dip}$. 
We estimate  Floquet quasi-energies by considering only the diagonals. We obtain:
\begin{equation}
\epsilon_l= \frac{1}{2}(A_i-A_j-A_k)(P_u + P_d) +C_{ij}+C_{ik}-C_{jk}
\label{eq:secular}
\end{equation}
where $i,j,k \equiv 1,2,3$ or cyclic permutations give $\epsilon_{l=1,2,3}$ quasienergies.

Thus we estimate the dip positions from the fact that the quasi-energies represent
the {\em gradients} of the spectral lines in Fig.\ref{Fig5} (i) and (ii), hence estimate the degeneracy point :
\begin{equation}
\tau^{(lm)}_{dip}\simeq \frac{2\pi}{\epsilon_l-\epsilon_m}
\label{eq:secdip}
\end{equation}
for the dip arising from the difference between the $l$ and $m$-th quasienergy.
One finding is that the  secular contribution from the dipolar coupling greatly amplifies the
effect of the (usually weaker) $C_{ij}$ dipolar coupling between the nuclei, as it is a linear contribution. This is in contrast to disjoint pairs; 
if the dipolar coupling is weak,
since $\omega_i = \pm \frac{1}{4} \sqrt{C_{12}^2+(P_i \Delta A)^2}$, for $C_{12} \ll  P_i \Delta A$
the non-secular contributions in the disjoint pairs represent a very small quadratic shift.

In terms of the interaction strengths, the two dips of the first doublet correspond to:
\begin{equation}
\tau^{(\pm 12)}_{dip}\simeq \frac{2\pi}{|\Delta_{12} (P_u + P_d) \pm 2(C_{31}-C_{23})|  }
\label{eq:secdip2}
\end{equation}
and similarly for other doublets. 

In Fig.\ref{Fig6} we compare values from Eq.\ref{eq:secdip2} with the full numerics.
Thus the mean position exposes the
 value of $\Delta_{12}$ while the splittings
expose the dipolar coefficients.

\section{ Conclusions}
The extension of technologies such as MRI and NMR to the nano-scale is an outstanding technical challenge which is
leading not only rapid experimental progress, but also the development of new methods to analyse data
and to optimise information gathering on the atomic scale structure. 

 Motivated by this, in the present work we introduce Floquet spectroscopy as an insightful new paradigm for understanding and analysis
of spin sensing experiments.  The approach is universally valid for any type of periodic driving whether
resonant or not. Hence, here potential applications have been explored for analysis of different physical regimes and sensors
 which are not necessarily associated with a single sharp resonant `dip' but may nevertheless potentially still offer well delineated  features. 

Our key findings are (i) that there is an underlying structure associated with Floquet avoided crossings and the Floquet spectrum which 
is potentially information rich; it represents an envelope on the usually studied coherence dips with a shape controlled by the widths of the 
avoided crossings. (ii) that the Floquet approach clarifies also regimes where the commonly-used average Hamiltonian theory methods will fail.
(iii) The method's generality extends beyond single spin and pseudospin systems and is also useful 
 for higher dimensional spin systems, and potentially any dynamical decoupling protocol, provided it is temporally periodic.

{\bf Acknowledgements} We are very grateful to Setrak Balian, Gary Wolfowicz and Gavin Morley for helpful discussions.

\appendix

\section{Floquet spectrum} 
\label{App:AppendixA}
\subsection{Coherence minima and avoided crossings}
A key result of the present work is that the coherence dips associated with single spin sensing 
are associated with avoided crossings of the underlying Floquet spectrum and in this appendix, 
this conjecture is justified.

In our study, we consider the important class of spin-sensing experiments for which an electronic sensor spin $S$ is coupled to each environmental  nucleus via the effective Hamiltonian:
\begin{equation}
\hat{H}\approx \langle i| \hat{S}_z | i \rangle  \bf{A} \cdot \bf{I}
\end{equation}
Where $i=u,d$ and $\bf{A}$ is a vector representing the hyperfine interaction. The dependence on $S_z$ only
arises because of the large energy difference between electronic and nuclear states; 
in the case of NV-centers, the above is valid only for magnetic fields of  magnitude and 
orientation which does not mix the electronic states. The result is a state conditional 
Hamiltonian:
\begin{equation}
\hat{H} = \frac{1}{\sqrt{2}} [|u\rangle \langle u| \otimes H_u + |d\rangle \langle d| \otimes H_d]
\end{equation}
where the $H_i$ are the effective bath Hamiltonians discussed in Section \ref{onespin} in the main text
so that an initial joint sensor-target spin state $\psi(0) \simeq \frac{1}{\sqrt{2}} (|u\rangle+|d\rangle) {\mathcal{B}}(0)$
evolves into an (in general) entangled state:
\begin{equation}
\psi(t) = \frac{1}{\sqrt{2}} (|u\rangle{\mathcal{B}}_u(t)+|d\rangle{\mathcal{B}}_d(t)).
\end{equation} 

Experiments probe the coherence   ${\mathcal L}(t)= \langle S^+ \rangle$. While experimental comparison
involves averaging over thermal bath states  $Tr[\rho S^+]$, without loss of generality we consider
a pure state $\mathcal{L}=\langle \psi(t) | S^+ | \psi(t) \rangle \propto \langle\mathcal{B}_u(t) | \mathcal{B}_d(t) \rangle$.

Maximum entanglement occurs whenever  $\langle\mathcal{B}_u(t)| \mathcal{B}_d(t)\rangle = 0$. However, the general condition for a minimum or dip to be observed is in fact $|\mathcal{B}_u(t)\rangle = -|{\mathcal{B}}_d(t)\rangle$. This is regardless of the particular dynamical decoupling sequence applied. The key question for design of an experimental  pulse sequence (say, CPMG) is for which pulse interval $\tau$ and total pulse number number $N \equiv 2N_p$ will correspond to underlying quantum evolution 
\begin{equation}
\langle {\mathcal{B}}_u(t=4N_p\tau) |{\mathcal{B}}_d(t=4N_p\tau)\rangle=-1
\label{Lmin}
\end{equation} 
 and thus a minimum in the function $ \mathcal{L}$,
which to within an unimportant normalisation factor we take, $ \mathcal{L}=\langle\mathcal{B}_u(t) | \mathcal{B}_d(t) \rangle$ (we note that for decoherence experiments probing $ |\mathcal{L}|$, this
in fact corresponds to a {\em maximum} in the coherence). 

The Floquet approach is based on the premise that for  any periodically  driven quantum system, the Floquet states $\Phi_{j} $ fulfil the same role as eigenstates of a Hamiltonian in a time-independent system. Thus if the initial quantum state is projected into a Floquet basis, i.e. ${\mathcal{B}}(0) =\sum_l a_l \Phi_{l}$, then its temporal evolution is known for all time.

A central finding for the present work is that for the pulsed dynamical decoupling, the eigenspectrum
is independent of the sensor spin state 
thus $\mathcal{E}_l^{(u)} = \mathcal{E}_l^{(d)} \equiv \mathcal{E}_l(\tau)$ where
the Floquet eigenspectrum $\mathcal{E}_l(\tau)$ is a function of the experimentally chosen pulse interval $\tau$. 

 The above results are proved in the next section,  but  we 
use them now to explain why coherence minima are associated with avoided crossings.
Since the Floquet spectra are the same, if there is an avoided crossing and thus a near degeneracy,
$e^{\mathcal{E}_l(\tau)} \simeq e^{\mathcal{E}_{k}(\tau)}$ in the $u$ subspace of sensor states, there will simultaneously be an avoided  crossing in the lower $d$ subspace of sensor states.

\subsection{Avoided crossings for two-level system}

Although the eigenspectra are the same, in general the corresponding eigenstates or Floquet states are not.
$\Phi_{d}(\tau) \neq  \Phi_{u}(\tau)$ for arbitrary $\tau$. Hence the temporal evolution:
\begin{eqnarray}
{\mathcal{B}}_u(t)&=& a_{u+} \Phi_{u+} e^{-i N_p {\mathcal E}(\tau)} + a_{u-} \Phi_{u-} e^{+i N_p {\mathcal E}(\tau)}  \neq\nonumber\\
{\mathcal{B}}_d(t)&=& a_{d+} \Phi_{d+} e^{-i N_p {\mathcal E}(\tau)} + a_{d-} \Phi_{d-} e^{+i N_p {\mathcal E}(\tau)} 
\end{eqnarray}
and thus entanglement  with the sensor is established since the sensor-target ${\mathcal{B}}_{u,d}(t)$ state is no longer 
factorisable. 

One exception occurs for  $\tau=0$, where all the Floquet states reduce to the  unperturbed (thermal states). For a two-state system, without loss of generality,
$ \Phi_{u+}(\tau=0)= \Phi_{d+}(\tau=0)=|\uparrow\rangle$ 
or alternatively $ \Phi_{u-}(\tau=0)= \Phi_{d-}(\tau=0)|\downarrow\rangle$. 

Another, most interesting, exception is at a level crossing, where the eigenstates take the same form.
It is a textbook result for level crossings (also known as Landau-Zener transitions) that the
the unperturbed states are maximally mixed and become sums and differences of the unperturbed
states.
The implication for the present case, is that the Floquet states for {\em both upper and lower} 
state {\em must } coincide at approximately $\frac{1}{\sqrt{2}} (|\downarrow\rangle\pm |\uparrow\rangle)$.
This allows for two distinct possibilities:\\
(i) In the first case,
\begin{eqnarray}
 \Phi_{u+} & = &\Phi_{d+} \simeq \frac{1}{\sqrt{2}} (|\downarrow\rangle +|\uparrow\rangle) \  \textrm{and} \nonumber\\  
  \Phi_{u-} & = &  \Phi_{d-} \simeq \frac{1}{\sqrt{2}} (|\downarrow\rangle- |\uparrow\rangle)
\end{eqnarray}
This possibility is the trivial case where the Floquet states for upper and lower sensor state are identical. 
There is never any difference in the evolution, no entanglement and so  
 $\langle {\mathcal{B}}_u(t) |{\mathcal{B}}_d(t)\rangle=+1$.\\
(ii) In the second case,\\
\begin{eqnarray}
 \Phi_{u+} &=&\Phi_{d-} \simeq \frac{1}{\sqrt{2}} (|\downarrow\rangle +|\uparrow\rangle)  \ \textrm{and} \nonumber\\  
  \Phi_{u-} &=&  \Phi_{d+} \simeq \frac{1}{\sqrt{2}} (|\downarrow\rangle- |\uparrow\rangle)
\end{eqnarray}
In this case, $\langle {\mathcal{B}}_u | {\mathcal{B}}_d\rangle=\cos{2N_p {\mathcal E}(\tau)}$
which may for an appropriate choice of $N_p \approx \pi/2{\mathcal E}$ attain the minimal value for a dip
$\langle {\mathcal{B}}_u | {\mathcal{B}}_d\rangle=-1$.

Diagonalisation of the two dimensional unitary matrix is straightforward (see \cite{SuppInfo}) and it is clear its eigenvalues must be conjugate pairs $\lambda_\pm = e^{\pm i \mathcal{E}}$. For a two-level case, the avoided crossing condition is $\lambda_+ = \lambda_-$ and hence coherence dips must occur at $\mathcal{E}\simeq 0, \pi, 2\pi$ with case (i) occurring at $\mathcal{E}\simeq \pi$ and case (ii) occurring at $\mathcal{E}\simeq 0,2\pi$.

\subsection{Coherence minima for the general case}

For a spin cluster of arbitrary size, an initial pure state can be projected into the upper or lower Floquet basis:
\begin{equation}
|\mathcal{B}(0)\rangle = \sum_l^D \langle\Phi_{il}|\mathcal{B}(0)\rangle |\Phi_{il}\rangle
\end{equation}
where $|\Phi_{il}\rangle$ are Floquet eigenstates for upper $(i = u)$ and lower $(i = d)$ sensor state
respectively and $D$ is the dimension of the one-period unitary evolution operator. The state after a time $t = 4N_p\tau$ is then
\begin{equation}
|\mathcal{B}_i(t)\rangle = \sum_l^D e^{-iN_p\mathcal{E}_l}\langle\Phi_{il}|\mathcal{B}(0)\rangle |\Phi_{il}\rangle
\end{equation}
where $\mathcal{E}_l$ are the eigenphases.

Thus the coherence decay of the sensor spin ($ \mathcal{L}(t) = \langle\mathcal{B}_l(t)|\mathcal{B}_u(t)\rangle$) is given in a Floquet basis by:
 \begin{equation}
 \mathcal{L}(t=N_p4\tau) = \sum^D_{l,l'} e^{-i N_p ({\mathcal E}_l-{\mathcal E}_{l'}) } |a_l|^2 | \langle \Phi_{dl'} | \Phi_{ul} \rangle|^2,
 \label{CoherenceGeneral}
 \end{equation}
Here $|a_l|^2 = |\langle\Phi_{ul}|\mathcal{B}(0)\rangle|^2$.  

The target spins are in fact in a thermal ensemble, which given small nuclear energy scale are all equally likely. Thus we must average over the thermal bath states and calculate $\langle\mathcal{L}\rangle = (1/D)\sum_j^D \mathcal{L}_j$, where $\mathcal{L}_j$ is the coherence function evaluated for the bath intitially in the j-th thermal bath state, $|\mathcal{B}_j(0)\rangle$. Under the thermal average we find  $\sum_j \langle\Phi_{ul}|\mathcal{B}_j(0)\rangle \langle \mathcal{B}_j(0)| \Phi_{ul}\rangle =1$. This produces Eq.~\ref{Fdeco} in the main text.
While  $\mathcal{L}$ for a pure state is complex, the (in any case small) imaginary part vanishes under the bath average
and we consider  only the real part of $\mathcal{L}$. 

Returning briefly to simplest case of $D = 2$, treated in the previous subsection,   Eq. (\ref{CoherenceGeneral}) may be rewritten as
\begin{equation}
 \langle\mathcal{L}(\tau)\rangle = 1 - 2|\langle\Phi_{d-}|\Phi_{u+}\rangle|^2\sin^2\left(N_p\frac{\mathcal{E}_1 - \mathcal{E}_2}{2}\right)
\end{equation}
which is of the form $\mathcal{L}(\tau) = 1 - F(\tau)f(N_p,\tau)$ where $f(N_p,\tau) \in [0,1]$ is a pulse number dependent oscillation. 
If we disregard the oscillation,  we obtain the pulse number independent minimal bound of the coherence function given by  $\mathcal{L}_\text{env}(\tau) = 1 - F(\tau)$, which we call the coherence \emph{envelope}.

 Even if  these envelopes are not necessarily sharp ``dips'' (especially in strong-coupling regimes) they can
correspond to sharply delineated structures (for both NV centers and donors) which
should still be observable experimentally and can yield valuable information about the atomic-scale structure. We have shown above that the dips occur when $|\Phi_{u+}\rangle = |\Phi_{d-}\rangle$. 

For a general bath, of arbitrary dimension $D$,  we  can re-arrange Eq.(\ref{CoherenceGeneral}) using only
orthonormality and completeness of the eigenstates  $\sum_{k'}|\langle\Phi_{dk'}|\phi\rangle|^2 = 1$ into paired contributions. 

\begin{eqnarray}
{\langle\mathcal{L}\rangle } \equiv  1 - \sum\limits_{l < l'} [| \langle \Phi_{dl'} | \Phi_{ul} \rangle|^2 + | \langle \Phi_{dl} | \Phi_{ul'} \rangle|^2 ]\nonumber\\ 
\times \sin^2(N_p\frac{\mathcal{E}_l - \mathcal{E}_{l'}}{2}) 
\end{eqnarray}

This is, again, composed of pulse number independent envelopes superimposed with pulse number dependent oscillations. For a minimum, we require the term in square brackets to be maximised; this will occur at a level crossing between a
given pair of eigenstates $l$ and $l'$ as argued in the previous subsection. For $D>2$, level crossings between $ \mathcal{E}_l \approx \mathcal{E}_{l'}$ occur at 
 arbitrary $\mathcal{E}_{l}$ and no longer at $ \mathcal{E}_{l,l'} \approx \pi$.
Dips occurring at the point for which $ | \langle \Phi_{dl'} | \Phi_{ul} \rangle|^2=1$ and
$ | \langle \Phi_{dl} | \Phi_{ul'} \rangle|^2 = 1$, generalises, to arbitrary dimension, the two-state orthogonality condition that $|\Phi_{u+}\rangle = |\Phi_{d-}\rangle$.\\

\newpage

\section{Symmetry of eigenphases for CPMG control}
For decoupling sequences like CPMG, the Floquet phases are independent of the sensor spin state, regardless of the dimensionality of the bath states i.e. $\mathcal{E}_l^{(u)} = \mathcal{E}_l^{(d)} \equiv \mathcal{E}_l$.  To show this we first construct the basic propagator, for total period $\tau_{tot}=4\tau$, which is to be repeated periodically:
\begin{eqnarray}
\hat{T}_{(u)}^{(2)}(4\tau) = \hat{T}_{(u)}(\tau) \hat{T}_{(d)}(\tau)\hat{T}_{(d)}(\tau) \hat{T}_{(u)}(\tau)\equiv \hat{T}_{(u)}(2\tau) \hat{T}_{(d)} (2\tau)  \nonumber \\
\hat{T}_{(d)}^{(2)}(4\tau)= \hat{T}_{(d)}(\tau) \hat{T}_{(u)}(\tau)\hat{T}_{(u)}(\tau) \hat{T}_{(d)}(\tau)\equiv \hat{T}_{(d)}(2\tau) \hat{T}_{(u)} (2\tau)  \nonumber\\
\label{prod}
\end{eqnarray}
We can then obtain the eigenvalues for $\hat{T}_{(u)}^{(2)}$:
\begin{equation}
\hat{T}_{(u)}^{(2)}(4\tau) |\Phi_{ul}\rangle= \hat{T}_{(u)}(2\tau)\hat{T}_{(d)}(2\tau) |\Phi_{ul}\rangle = e^{-i\mathcal{E}_l}|\Phi_{ul}\rangle.
\end{equation}
Here $\exp(-i\mathcal{E}_l)$ is the $l$th eigenvalue of $\hat{T}_{(u)}^{(2)}$. If we apply the half period operator, 
$\hat{T}_{(d)}(2\tau)$:
\begin{equation}
\hat{T}_{(d)}(2\tau)\hat{T}_{(u)}(2\tau)\hat{T}_{(d)}(2\tau)|\Phi_{ul}\rangle = e^{-i\mathcal{E}_l}\hat{T}_{(d)}(2\tau)|\Phi_{ul}\rangle,
\end{equation}
this is equivalent to
\begin{equation}
\hat{T}_{(d)}^{(2)}(4\tau) \hat{T}_{(d)}(2\tau)|\Phi_{ul}\rangle = e^{-i\mathcal{E}_l}\hat{T}_{(d)}(2\tau)|\Phi_{ul}\rangle,
\label{F State Swaps}
\end{equation}
Thus $\exp(-i\mathcal{E}_l)$ is also an eigenvalue of $\hat{T}_{(d)}^{(2)}$. Eq.\ref{F State Swaps}
 implies that $\hat{T}_{(d)}(2\tau)|\Phi_{ul}\rangle$ is a eigenstate of $\hat{T}_{(d)}^{(2)}$, i.e. 
 $\hat{T}_{(d)}(2\tau) |\Phi_{ul}\rangle \propto |\Phi_{dl}\rangle$, where the factor of proportionality is a complex phase
$\exp(i \mu_{ld})$. Similarly,  
\begin{equation}
\hat{T}_{(u)}(2\tau)|\Phi_{dl}\rangle = \exp(i \mu_{lu}) |\Phi_{ul}\rangle
\end{equation}
for which $\mu_{ld}+\mu_{lu}= \mathcal{E}_l$.
From this we see that each Floquet state $|\Phi_{ul}\rangle, |\Phi_{dl}\rangle$ is the half-period evolution of the other (up to a complex phase).

\section{Pulses of finite duration}

Provided the $\hat{T}_{(u,d)}^{(2)}(4\tau)$ can be decomposed into products of sub-propagators, as in Eq.\ref{prod},
the $\pi$ pulses do not have to be of very short duration. If we write:
\begin{eqnarray}
 \hat{T}_{(u)}(2\tau+2\delta) = \hat{T}_{(u)}(\tau) T_{\pi}(2\delta) \hat{T}_{(d)}(\tau) \nonumber \\
\hat{T}_{(d)}(2\tau+2\delta) = \hat{T}_{(d)}(\tau) T_{\pi}(2\delta) \hat{T}_{(u)}(\tau)
\label{finite}
\end{eqnarray}
Then we can write the full propagator in the same form as previously:

 \begin{eqnarray}
\hat{T}_{(u)}^{(2)}(4\tau') &=& \hat{T}_{(u)}(2\tau') \hat{T}_{(d)} (2\tau')  \nonumber \\
\hat{T}_{(d)}^{(2)}(4\tau')&=&  \hat{T}_{(d)}(2\tau') \hat{T}_{(u)} (2\tau')  
\label{prod}
\end{eqnarray}

but with $\tau'=\tau+\delta$. Then all the above follow; the pulses of finite duration are still assumed to be $\pi$ pulses
but there can be some arbitrary evolution of the system during the finite interval $\delta$, but properties such as the state independence of the Floquet phases still hold.

\end{document}